\documentclass[showpacs,aps,
twocolumn,showpacs,preprintnumbers,superscriptaddress]{revtex4-1}
\usepackage{amsmath}
\usepackage{graphicx}
\usepackage{dcolumn}
\usepackage{bm}
\usepackage{color}
\usepackage{epstopdf}

\graphicspath{{figs/}}

\begin{document}
\title{Ramsey Interferometry based on stimulated Brillouin scattering}
\author{Xian-Fu Quan}
\affiliation{State Key Laboratory of Magnetic Resonance and Atomic and Molecular Physics,
Wuhan Institute of Physics and Mathematics, Chinese Academy of Sciences, Wuhan 430071, China}
\affiliation{University of Chinese Academy of Sciences, Beijing 100049, China}
\author{Zhi-Rui Gong}
\affiliation{The College of Physics and Energy, Shenzhen University, Shenzhen 518060, China }
\author{Shuo Zhang}
 \affiliation{Zhengzhou Information Science and Technology Institute, Zhengzhou 450004, China}
 \author{Liang Chen}
\affiliation{State Key Laboratory of Magnetic Resonance and Atomic and Molecular Physics,
Wuhan Institute of Physics and Mathematics, Chinese Academy of Sciences, Wuhan 430071, China}
\author{Jian-Qi Zhang}
\email{changjianqi@gmail.com}
\affiliation{State Key Laboratory of Magnetic Resonance and Atomic and Molecular Physics,
Wuhan Institute of Physics and Mathematics, Chinese Academy of Sciences, Wuhan 430071, China}
\author{Mang Feng}
\email{mangfeng@wipm.ac.cn}
\affiliation{State Key Laboratory of Magnetic Resonance and Atomic and Molecular Physics,
Wuhan Institute of Physics and Mathematics, Chinese Academy of Sciences, Wuhan 430071, China}
 \affiliation{Department of Physics, Zhejiang Normal University, Jinhua 321004, China}

\begin{abstract}
Realizing highly sensitive interferometry is essential to accurate observation of quantum properties. Here we study two kinds of Ramsey
interference fringes in a whispering-gallery resonator, where the coherent phonons for free evolution can be achieved by stimulated Brillouin scattering.
These two different fringes appear, respectively, in the regimes of rotating wave approximation (RWA) and anti-RWA. Our work shows particularly that the anti-RWA Ramsey interference takes some quantum properties of squeezing, which enhances the strength and
visibility of the fringes and shows robustness against the system's decay. In application, our proposal,
feasible with current laboratory techniques, provides a practical idea for building better quantum interferometers.
\end{abstract}

\pacs{42.50.-p, 03.75.Dg, 42.50.Nn}

\maketitle

\section{Introduction}

Precision measurement plays an essential role in modern physics. Ramsey interferometry \cite{rmp-62-541,pr-79-695}, based on
separated oscillatory fields, has been employed for precision measurement for several decades. This interferometry was first proposed to investigate the molecular beam
resonance based on Rabi's method \cite{pr-79-695}, and then widely applied to different
fields, such as atomic clocks \cite{prl-117-143004}, quantum information
processing \cite{prl-90-143602}, quantum simulation \cite{pra-77-033832} and
optimized quantum measurements \cite{naturephonoics-5-222}. However, all
these schemes are based on the Ramsey interference (RI) regarding internal states of
the atoms. As such, it is interesting to consider the Ramsey interferometry by other ideas.

The present work focuses on the Ramsey interferometry based on stimulated Brillouin
scattering (SBS) processes \cite{NonOpt}. The SBS is caused by
electrostriction and photon-elastic effects, which is a fundamental
interaction between photons and phonons in fibers and waveguides \cite
{nphys-11-275}. Recently, the SBS has presented considerable potential in a variety
of optical applications, such as light storage \cite{science-318-1748}, SBS induced
transparency \cite{nphys-11-275,ncomm-6-6193}, slow light \cite{prl-94-153902}
, laser linewidth narrowing \cite{JOSAB-18-556}, and optical isolators \cite
{JLT-29-2267}. Especially in photonic integrated circuits, the SBS
can be greatly enhanced by tight confinement of the optical fields \cite
{ncomm-4-1944}. On the other hand, we have noticed that the optomechanics can be
applied to construct the Ramsey interferometry using nanomechanical resonator vibration \cite
{pra-90-053809}. This reminds us of a question: Can the processes of
SBS, with analogy to optomechanics, be used to achieve the RI, even beyond ?

Here we propose a scheme to realize two kinds of RI
fringes with SBS in a whispering-gallery resonator (WGR).
These two different fringes work in the RWA and anti-RWA regimes,
respectively, which can be controlled by the driving fields on
different optical modes. Each of our RI is composed of three steps like the
previous idea: Creation of phonon coherent states in the acoustic mode, free
evolution of the phonons, and then quantum interference between the phonons and
photons. The relative phase for the interference is from free
evolution of the phonons due to its long coherent time.

Compared with the previously proposed RI \cite{pra-90-053809,quantumoptics}, our scheme
has some significantly different characters. Firstly, the fringes based on the anti-RWA can get benefits from squeezing
properties for better visibility, higher strength and more robustness to decay. Secondly, different from most
of the previous works with RI, where fermions, e.g., electrons, are employed, our scheme is based on the bosons (phonons and photons).
Thirdly, in comparison with \cite{pra-90-053809} where the relative phase is accumulated by the
external vibration of the optomechanical resonator, the relative phase in our scheme for the interference is
achieved via free evolution of the phonons in the acoustic mode. This is due to the fact that acoustic
phonons circulating along the surface of the WGR experience very small decay. Although \cite{pra-90-053809} provided an idea for RI with bosons,
our work, based on the SBS working in the anti-RWA regime, represents a new bosonic method to realize the RI.

The rest of our work is structured as follows. We present the model of the SBS in the next section, and solve analytically
the two kinds of RI fringes in Sec. III. Numerical simulations along with some discussion is given in Sec. IV. The last section
is for a brief conclusion.

\section{System and Model}
\begin{figure}[tbph]
\centering\includegraphics[width=6cm]{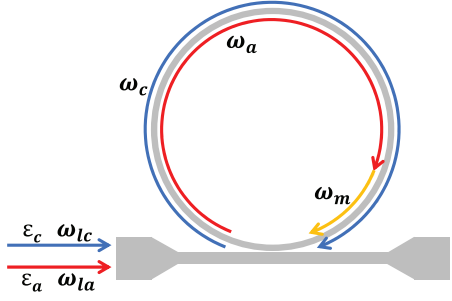}
\caption{Schematic diagram of the system. Both control and probe fields are coupled to
phase-matched optical modes of the whispering-gallery resonator by means of a
tapered optical fibre. A forward-SBS opto-acoustic interaction takes place
inside the resonator. Here the control and probe fields are determined by
the driving strengths, which also determines whether the work region is in RWA or anti-RWA
[see details in Sec. III].}
\label{model}
\end{figure}
We consider a WGR as plotted in Fig. \ref{model}, where both optical and acoustic waves
circulate along its surface \cite{nphys-11-275,ncomm-6-6193}. When the
acoustic mode $b$ and two optical modes $a$ and $c$ fulfill the energy and
momentum conservations, the photons are scattered between the two
bosons (i.e., acoustic and optical) modes via the SBS processes\cite{nphys-11-275,ncomm-6-6193,pra-88-013815}, which is expressed as%
\begin{equation}
H_{B}=g(abc^{\dag }+a^{\dag }b^{\dag }c),  \label{101}
\end{equation}%
where $g$ is the single photon Brillouin scattering coupling strength, $b$
is the annihilation operator for the acoustic mode, and $a$ and $c$ are the
annihilation operators for the optical modes. The optical mode $j$ is driven
by a time-dependent external field $\varepsilon_{j}$ with the frequency $%
\omega _{lj}$ ($j=a,c$). These processes can be described as below under the control and probe fields
\begin{equation}
\begin{array}{ccc}
H_{d} & = & i[\varepsilon _{c}(t)c^{\dag }e^{-i\omega _{lc}t}+\varepsilon
_{a}(t)a^{\dag }e^{-i\omega _{la}t}-H.c.],
\end{array}
\label{102}
\end{equation}%
where $\varepsilon _{j}(t)=\sqrt{\frac{2\kappa _{j}\wp _{j}(t)}{\hbar\omega_{lj}}}$
is the time-dependent driving strength. $\kappa_{j}$ and $\wp_{j}$
correspond, respectively, to the decay rate and the external field power regarding the
$j$th optical mode.

The whole system with the optical mode $a$ ($c$) and the acoustic mode $b$ at frequencies $\omega_{a}$ ($%
\omega_{c}$) and $\omega_{m}$, is governed by the Hamiltonian
\begin{equation}
\begin{array}{ccc}
H & = & \omega _{a}a^{\dag }a+\omega _{m}b^{\dag }b+\omega _{c}c^{\dag
}c+H_{B}+H_{d}.
\end{array}
\label{103}
\end{equation}

\section{Analytical solutions to the RI fringes}

\begin{figure}[tbph]
\centering
\includegraphics[width=8cm]{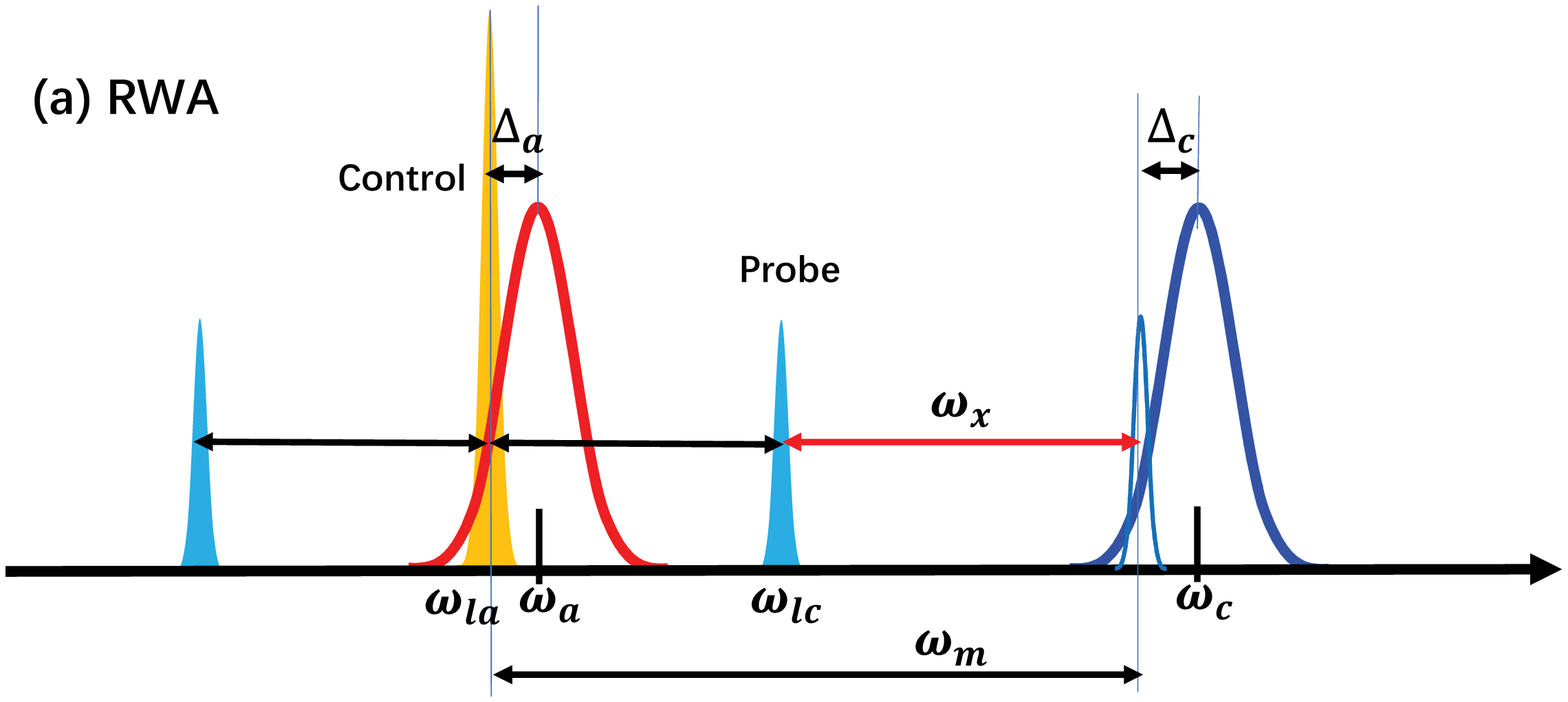} \includegraphics[width=8cm]{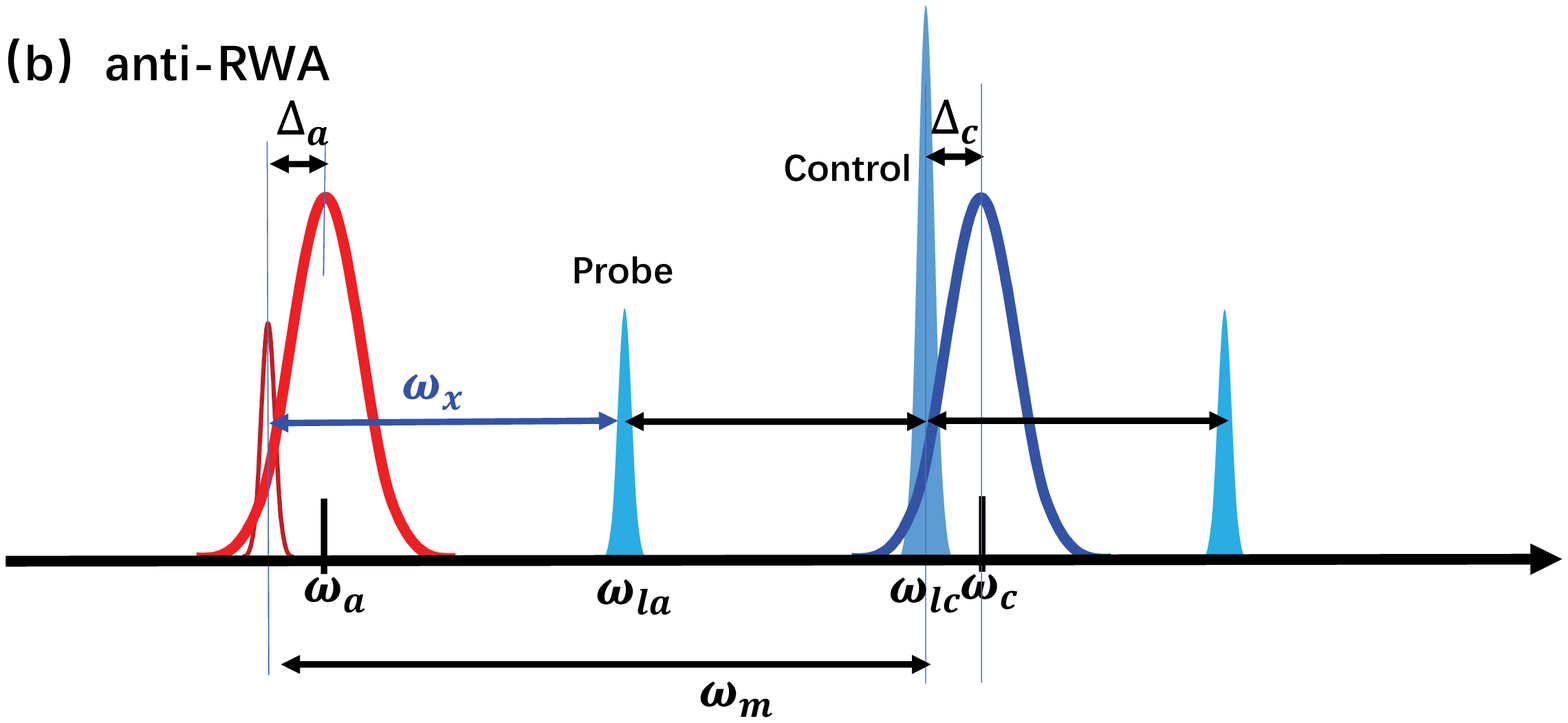}
\caption{(a) Frequency relationship of the coupled triplet system
for RI fringes in the region of RWA, where the control field on mode $a$
is of lower frequency than the anti-Stokes mode $c$. When the modulation
frequency sweeps over the fixed Brillouin phase match frequency, $\protect%
\omega_{m}$, the RI fringes can be observed in the probe field at the
anti-Stokes mode $c$. (b) Frequency relationship of the coupled triplet system
for RI fringes in the region of anti-RWA, where the control field on the
mode $c$ is of higher frequency than the probe field on the Stokes mode $a$, and
the RI fringes can be detected in the optical mode $a$.}
\label{ORAB}
\end{figure}

With employment of Eq. (\ref{103}), we may derive solutions analytically for RI fringes under the conditions of RWA and
anti-RWA. The corresponding frequency relationships for the control
field, the probe field and the acoustic mode are presented in Fig.\ref{ORAB}.

\subsection{RI fringes in the regime of RWA}

\begin{figure}[tbph]
\centering\includegraphics[width=6cm]{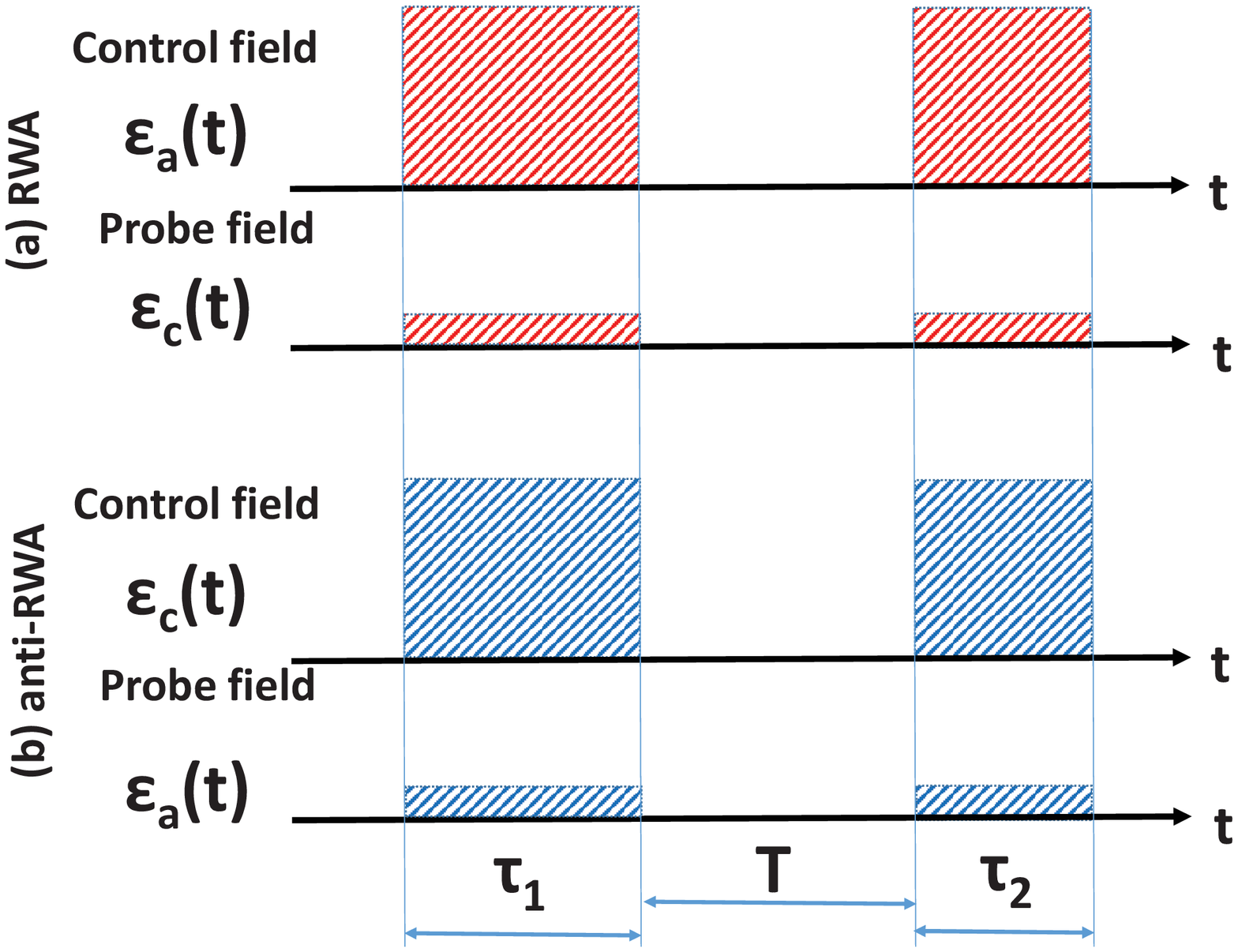}
\caption{Sketch of pulses of control and probe fields for the RI fringes in the regimes of RWA (a) and anti-RWA (b)}
\label{square}
\end{figure}

When the driving on mode $a$ is much stronger than that on mode $c$, i.e., $\varepsilon _{a}(t)\gg \varepsilon _{c}(t)$ (see Fig.\ref{ORAB}(a)), under
the condition of parametric approximation, we can treat the control mode
$a$ as a complex number. With the application of non-depletion
approximation for the nonlinear three-wave mixing, we assume that mode $a$
is not influenced by SBS, and thereby the dynamics of mode $a$ is governed by%
\begin{equation}
\begin{array}{ccl}
\dfrac{d}{dt}a & = & -i\omega _{a}a-\kappa _{a}a+\varepsilon
_{a}(t)e^{-i\omega _{la}t}.
\end{array}
\end{equation}
The steady state solution of mode $a$ is $\alpha (t)=\frac{
\varepsilon _{a}(t)}{\kappa _{a}+i(\omega _{a}-\omega _{la})}$ with the square pulses [see Fig.\ref{square}(a)]
\begin{equation}
\varepsilon _{a}(t)=\left\{
\begin{array}{cc}
\varepsilon _{a} & (0\leq t\leq \tau _{1}) \\
0 & (\tau _{1}\leq t\leq \tau _{1}+T) \\
\varepsilon _{a} & (\tau _{1}+T\leq t\leq \tau _{1}+T+\tau _{2}).
\end{array}
\right.
\end{equation}

By inserting
this complex number to Eq. (\ref{103}), the linearized Hamiltonian for photon-phonon interaction is
\begin{equation}
\begin{array}{ccl}
H_{R} & = & \omega _{m}b^{\dag }b+\omega _{c}c^{\dag }c+i\varepsilon
_{c}(t)(c^{\dag }e^{-i\omega _{lc}t}-ce^{i\omega _{lc}t}) \\
& + & g[\alpha (t)bc^{\dag }e^{-i\omega _{la}t}+\alpha ^{\ast }(t)b^{\dag
}ce^{i\omega _{la}t}].
\end{array}
\label{1031}
\end{equation}
Within the rotating frame of $H_{r}=\omega _{lc}c^{\dag }c+(\omega _{lc}-\omega
_{la})b^{\dag }b$, the above Hamiltonian can be reduced to a RWA Hamiltonian
\begin{equation}
\begin{array}{ccl}
H_{RWA} & = & \omega _{x}b^{\dag }b+\Delta _{c}c^{\dag }c+g[\alpha
(t)bc^{\dag }+\alpha ^{\ast }(t)b^{\dag }c] \\
& + & i\varepsilon _{c}(t)(c^{\dag }-c),
\end{array}
\label{104}
\end{equation}%
with $\Delta _{c}=\omega _{c}-\omega_{lc}$ and $\omega _{x}=\omega
_{m}-(\omega _{lc}-\omega _{la})$ \cite{nphys-11-275}. Note that
the probe field $\varepsilon _{c}(t)$ is denoted by time-dependent squared pulses, as plotted in Fig. \ref{square}(a).

With employment of Heisenberg-Langevin equations and the mean-field
approximation, the mean response of the system for the weak field $%
\varepsilon_{c}(t)$ is governed by the mean-value equations as
\begin{equation}
\begin{array}{ccl}
\dfrac{d}{dt}c & = & -i\Delta _{c}c-iG_{r}(t)b-\kappa _{c}c+\varepsilon
_{c}(t), \\
\dfrac{d}{dt}b & = & -i\omega _{x}b-iG_{r}^{\ast }(t)c-\dfrac{\gamma _{m}}{2}%
b,%
\end{array}
\label{201}
\end{equation}%
where $G_{r}(t)=g\alpha (t)$ is only controlled by the field $%
\varepsilon _{a}(t)$, while the driving field $\varepsilon _{c}(t)$ on mode $c$
works as a probe field.

In this case, the output field for the probe field $\varepsilon _{c}(t)$
at frequency $\omega _{lc}$ can be described by the input-output relation $%
\varepsilon_{c-out}(t)=2\kappa _{c}c-\varepsilon _{c}(t)$ at any time $t$.
Using the Fourier transform $\left\langle O(\omega
)\right\rangle =\int O(t)e^{-i\omega t}dt$, we have
\begin{equation}
\begin{array}{ccl}
-i\omega c & = & -i\Delta _{c}c-iG_{r}(t)b-\kappa _{c}c+\varepsilon
_{c}\delta _{\omega ,0}, \\
-i\omega b & = & -i\omega _{x}b-iG_{r}^{\ast }(t)c-\dfrac{\gamma _{m}}{2}b,%
\end{array}
\label{202}
\end{equation}
with the corresponding solutions as
\begin{equation}
\begin{array}{ccc}
c & = & \dfrac{\varepsilon _{c}\delta _{\omega ,0}[\dfrac{\gamma _{m}}{2}%
+i(\omega _{x}-\omega )]}{[\kappa _{c}+i(\Delta _{c}-\omega )][\dfrac{\gamma
_{m}}{2}+i(\omega _{x}-\omega )]+|G_{r}|^{2}}, \\
b & = & \dfrac{-iG_{r}^{\ast }\varepsilon _{c}\delta _{\omega ,0}}{[\kappa
_{c}+i(\Delta _{c}-\omega )][\dfrac{\gamma _{m}}{2}+i(\omega _{x}-\omega
)]+|G_{r}|^{2}}.%
\end{array}
\label{203}
\end{equation}

To get the solutions analytically, we suppose $\Delta _{c}\simeq\omega$ and $%
\Gamma_{r}=\dfrac{\gamma _{m}}{2}+\dfrac{|G_{r}|^{2}}{\kappa _{c}}$. Thus
the above equation (\ref{203}) is reduced to
\begin{equation}
\begin{array}{ccl}
\kappa _{c}c & \simeq & \dfrac{\varepsilon _{c}\delta _{\omega ,0}[\dfrac{%
\gamma _{m}}{2}+i(\omega _{x}-\omega )]}{[\Gamma _{r}+i(\omega _{x}-\omega )]%
}, \\
iG_{r}b & \simeq & \dfrac{|G_{r}|^{2}\varepsilon _{c}\delta _{\omega ,0}}{%
\kappa _{c}[\Gamma _{r}+i(\omega _{x}-\omega )]},%
\end{array}
\label{204}
\end{equation}
which can be rewritten as%
\begin{equation}
\begin{array}{rcc}
\kappa _{c}c+iG_{r}b & \simeq & \varepsilon _{c}\delta _{\omega ,0}, \\
\lbrack \Gamma _{r}+i(\omega _{x}-\omega )]b & \simeq & \dfrac{-iG_{r}^{\ast
}\varepsilon _{c}\delta _{\omega ,0}}{\kappa _{c}}.
\end{array}
\label{205}
\end{equation}
Using the Fourier inverse transformation on above equations, we obtain
\begin{equation}
\begin{array}{rcc}
\kappa _{c}c+iG_{r}b & \simeq & \varepsilon _{c}, \\
\dfrac{d}{dt}b+(\Gamma _{r}+i\omega _{x})b & \simeq & \dfrac{-iG_{r}^{\ast
}\varepsilon _{c}}{\kappa _{c}}.%
\end{array}
\label{206}
\end{equation}

As a result, the RI with RWA can be realized by following three steps.

Step 1 -- First pulse: Generation of the coherent acoustic mode $b$

With the initial condition $b(0)=0$, under the action of the driving field $%
\varepsilon _{c}$ for a time interval $\tau _{1}$, the dynamics of the
acoustic mode is given by
\begin{equation}
\begin{array}{ccc}
\kappa _{c}b & \simeq & -iG_{r}^{\ast }\varepsilon _{c}\dfrac{1-\exp
[-(\Gamma _{r}+i\omega _{x})\tau _{1}]}{\Gamma _{r}+i\omega _{x}},%
\end{array}
\label{207}
\end{equation}%
corresponding to the first $\pi /2$ pulse in the atom RI.

In this step, under the driving of the field $\varepsilon _{a}$, the
anti-Stokes processes for RWA dominate the system. As such, the energy is
transferred between the optical mode $c$ and the acoustic mode $b$, and thus the
acoustic mode is coherent.

Step 2 -- Free evolution of the acoustic mode $b$

In the time interval $[\tau _{1},\tau _{1}+T]$, the dynamics for the optical
and acoustic modes is governed by
\begin{equation}
\begin{array}{ccl}
\dfrac{d}{dt}c & = & -i\Delta _{c}c-\kappa _{c}c+\varepsilon _{c}(t), \\
\dfrac{d}{dt}b & = & -i\omega _{x}b-\dfrac{\gamma _{m}}{2}b.
\end{array}  \label{208}
\end{equation}%
Combined with the final solution for the first pulse, the dynamics of
acoustic mode turns to be
\begin{equation}
\begin{array}{ccc}
\kappa _{\mathrm{c}}b & \simeq & -iG_{r}^{\ast }\varepsilon _{\mathrm{c}}%
\dfrac{1-\exp [-(\Gamma _{r}+i\omega _{x})\tau _{1}]}{\Gamma _{r}+i\omega
_{x}}\exp [-(i\omega _{x}+\dfrac{\gamma _{m}}{2})T],%
\end{array}
\label{209}
\end{equation}
where the phase for interference, i.e., $\exp [-(i\omega _{m}+\dfrac{\gamma _{m}}{2}%
)T]$, is from free dynamics of the acoustic mode.

In this step, it is necessary to ensure that the decay rate of the optical
modes is large enough so that the acoustic mode $b$ is decoupled from
the optical mode $c$ very quickly, that is, $exp[\kappa _{\mathrm{c}}t^{\prime
}]=0$ for $0<t^{\prime}\ll T$. As such,
the acoustic mode $b$ can get the relative phase from free evolution
with a very small decay, i.e., $exp[\dfrac{\gamma _{m}}{2}T]\simeq 1$.

\begin{figure*}[tbph]
\centering
\includegraphics[width=16cm]{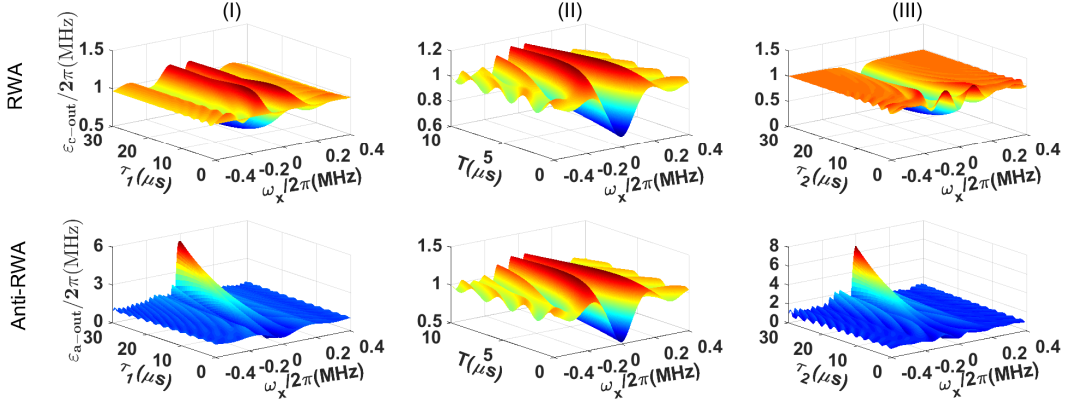}
\caption{(I) RI fringes work as a function of the operation time $%
\protect\tau _{1}$ ($T=4\protect\mu $s and $\protect\tau _{2}=0.1\protect\mu
$s); (II) RI fringes versus the free evolution time $T$ ($\protect%
\tau _{1}=4\protect\mu $s and $\protect\tau _{2}=0.1$ $\protect\mu$s); (III)
RI fringes with the different operation time $\protect\tau _{2}$ ($%
\protect\tau _{1}=4$ $\protect\mu$s and $T=4$ $\protect\mu$s). Here the Ramsey
fringes are inverse to the conventional ones as in Ref. \protect\cite%
{pra-90-053809}. Other parameters are from Refs. \protect\cite%
{nphys-11-275,ncomm-6-6193,pra-90-053809}: $\protect\omega _{m}=42.3$ MHz, $\protect\varepsilon _{c}=$ $2\protect\pi\times 1$ MHz (RWA), $%
\protect\varepsilon _{a}=$ $2\protect\pi\times 1$ MHz (anti-RWA), $%
|G_{r}|=|G_{b}|=2\protect\pi \times 0.58$ MHz, $\protect\kappa _{a}=\protect%
\kappa _{c}=2\protect\pi \times 40$ MHz, $\protect\gamma_{m}=2\protect%
\pi \times 20$ kHz and $\protect\omega _{lc}-\protect\omega _{la}\simeq
\protect\omega_{m}$.}
\label{fig33}
\end{figure*}

Step 3 -- Second pulse: The solution of the RI fringes

With application of the second pulse in the time interval $[\tau
_{1}+T,\tau _{1}+T+\tau _{2}]$, the solution for the acoustic RI fringes
following Eq.(\ref{201}) is
\begin{equation}
\begin{array}{ccl}
\kappa _{c}b & \simeq & -iG_{r}^{\ast }\varepsilon _{c}\dfrac{1-\exp
[-(\Gamma _{r}+i\omega _{x})\tau _{1}]}{\Gamma _{r}+i\omega _{x}}\exp
[-i\phi -\theta _{r}] \\
&- & iG_{r}^{\ast }\varepsilon _{c}\dfrac{1-\exp [-(\Gamma _{r}+i\omega
_{x})\tau _{2}]}{(\Gamma _{r}+i\omega _{x})}%
\end{array},
\end{equation}
with $\phi =\omega _{x}(T+\tau _{2})$ and $\theta _{r}=\dfrac{\gamma _{m}}{2}%
T+\Gamma _{r}\tau _{2}$. The corresponding solution for the optical mode $c$
is
\begin{equation}
\begin{array}{ccl}
c & \simeq  & \dfrac{\varepsilon _{c}}{\kappa _{c}}\{1+\dfrac{|G_{r}|^{2}}{%
\kappa _{c}}(\dfrac{\exp [-(\Gamma _{r}+i\omega _{x})\tau _{1}]-1}{\Gamma
_{r}+i\omega _{x}}\exp [-i\phi -\theta _{r}] \\
& + & \dfrac{\exp [-(\Gamma _{r}+i\omega _{x})\tau _{2}]-1}{(\Gamma
_{r}+i\omega _{x})})\},
\end{array}
\label{33c}
\end{equation}%
for the output field $\varepsilon _{c-out}=2\kappa _{c}c-\varepsilon _{c}$.

\subsection{RI fringes in the regime of anti-RWA}

Similar to the RI fringes in the RWA regime, under the parametric approximation, when the driving on
mode $c$ is much stronger than that on mode $a$, i,e, $\varepsilon_{c}(t)\gg\varepsilon_{a}(t)$
[see Fig.\ref{ORAB}(b)], the Hamiltonian in Eq. (\ref{103}), in the rotating frame $H_{b}=\omega _{la}a^{\dag
}a+(\omega _{la}-\omega _{lc})b^{\dag }b$, can be reduced to an anti-RWA Hamiltonian,
\begin{equation}
\begin{array}{ccl}
H_{NRWA} & = & \omega _{x}b^{\dag }b+\Delta _{a}a^{\dag }a+g(\zeta ^{\ast
}ab+\zeta a^{\dag }b^{\dag }) \\
& + & i\varepsilon _{a}(a^{\dag }-a),
\end{array}
\label{105}
\end{equation}%
where $\Delta _{a}=\omega _{a}-\omega _{la}$ and $\zeta =\frac{\varepsilon _{c}(t)}{\kappa _{a}+i(\omega _{c}-\omega _{lc})}
$  with a time-dependent
squared pulses [see Fig.\ref{square} (b)]
\begin{equation}
\varepsilon _{c}(t)=\left\{
\begin{array}{cc}
\varepsilon _{c} & (0\leq t\leq \tau _{1}) \\
0 & (\tau _{1}\leq t\leq \tau _{1}+T) \\
\varepsilon _{c} & (\tau _{1}+T\leq t\leq \tau _{1}+T+\tau _{2}).
\end{array}%
\right.
\end{equation}
The corresponding Langevin
equations are given by
\begin{equation}
\begin{array}{ccl}
\dfrac{d}{dt}a & = & -i\Delta _{a}a-iG_{b}(t)b^{\dag }-\kappa
_{a}a+\varepsilon _{a}(t), \\
\dfrac{d}{dt}b^{\dag } & = & i\omega _{x}b^{\dag }+iG_{b}^{\ast }(t)a-\dfrac{%
\gamma _{m}}{2}b^{\dag },%
\end{array}
\label{301}
\end{equation}%
where $G_{b}(t)=g\zeta (t)$ is driven by the field $\zeta
(t) $.

As a result, we obtain the solutions of the RI fringes for anti-RWA as
\begin{equation}
\begin{array}{ccl}
\kappa _{a}a & \simeq  & \varepsilon _{a}-iG_{b}(t)b^{\dag}, \\
\kappa _{a}b & \simeq  & iG_{b}\varepsilon _{a}\dfrac{\exp [-(\Gamma
_{b}+i\omega _{x})\tau _{1}]-1}{\Gamma _{b}+i\omega _{x}}\exp [-i\phi
-\theta _{b}] \\
& + & iG_{b}\varepsilon _{a}\dfrac{\exp [-(\Gamma _{b}+i\omega _{x})\tau
_{2}]-1}{(\Gamma _{b}+i\omega _{x})},
\end{array}
\label{302}
\end{equation}%
with $\Gamma _{b}=\dfrac{\gamma _{m}}{2}-\dfrac{|G_{b}|^{2}}{\kappa _{a}}$
and $\theta _{b}=\dfrac{\gamma _{m}}{2}T+\Gamma _{b}\tau _{2}$. The solution
for the optical mode $a$ is%
\begin{equation}
\begin{array}{ccl}
a & \simeq  & \dfrac{\varepsilon _{a}}{\kappa _{a}}\{1+\dfrac{|G_{b}|^{2}}{%
\kappa _{a}}(\dfrac{\exp [-(\Gamma _{b}-i\omega _{x})\tau _{1}]-1}{\Gamma
_{b}-i\omega _{x}}\exp [i\phi -\theta _{b}]) \\
& + & \dfrac{\exp [-(\Gamma _{b}-i\omega _{x})\tau _{2}]-1}{(\Gamma
_{b}-i\omega _{x})}\},
\end{array}
\label{33a}
\end{equation}%
for the output field, i.e., $\varepsilon _{a-out}=2\kappa _{a}a-\varepsilon _{a}$.

\section{Simulations and Discussion}

To further clarify the RI based on the SBS processes, we present below some
numerical simulations and discuss the RI fringes based on Eqs. (\ref{33c})
and (\ref{33a}) for optical modes $c$ and $a$ as well as their physical
properties, especially for mode $a$ in the anti-RWA regime.

\begin{figure}[tbph]
\centering
\includegraphics[width=8cm]{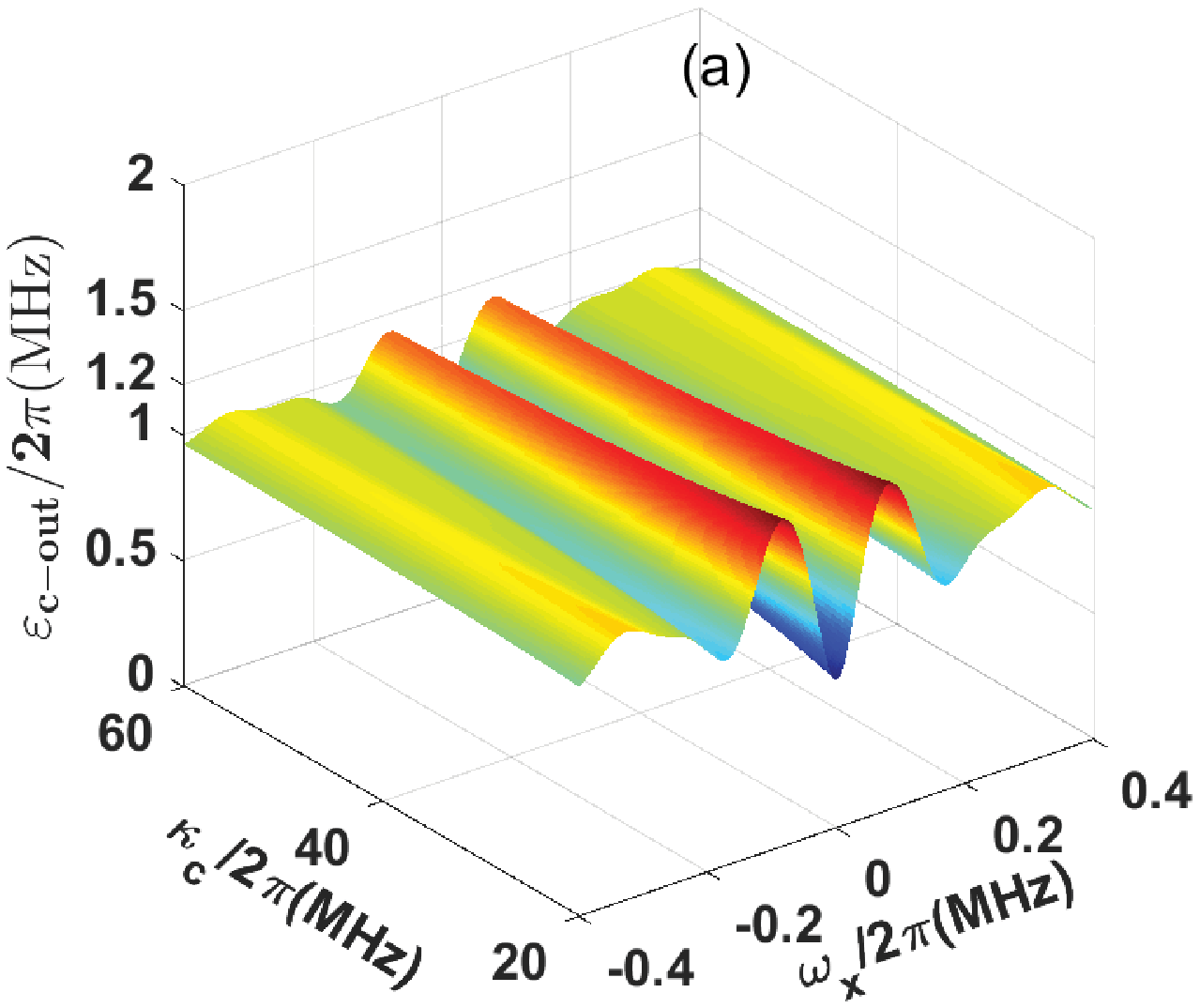}
\includegraphics[width=8cm]{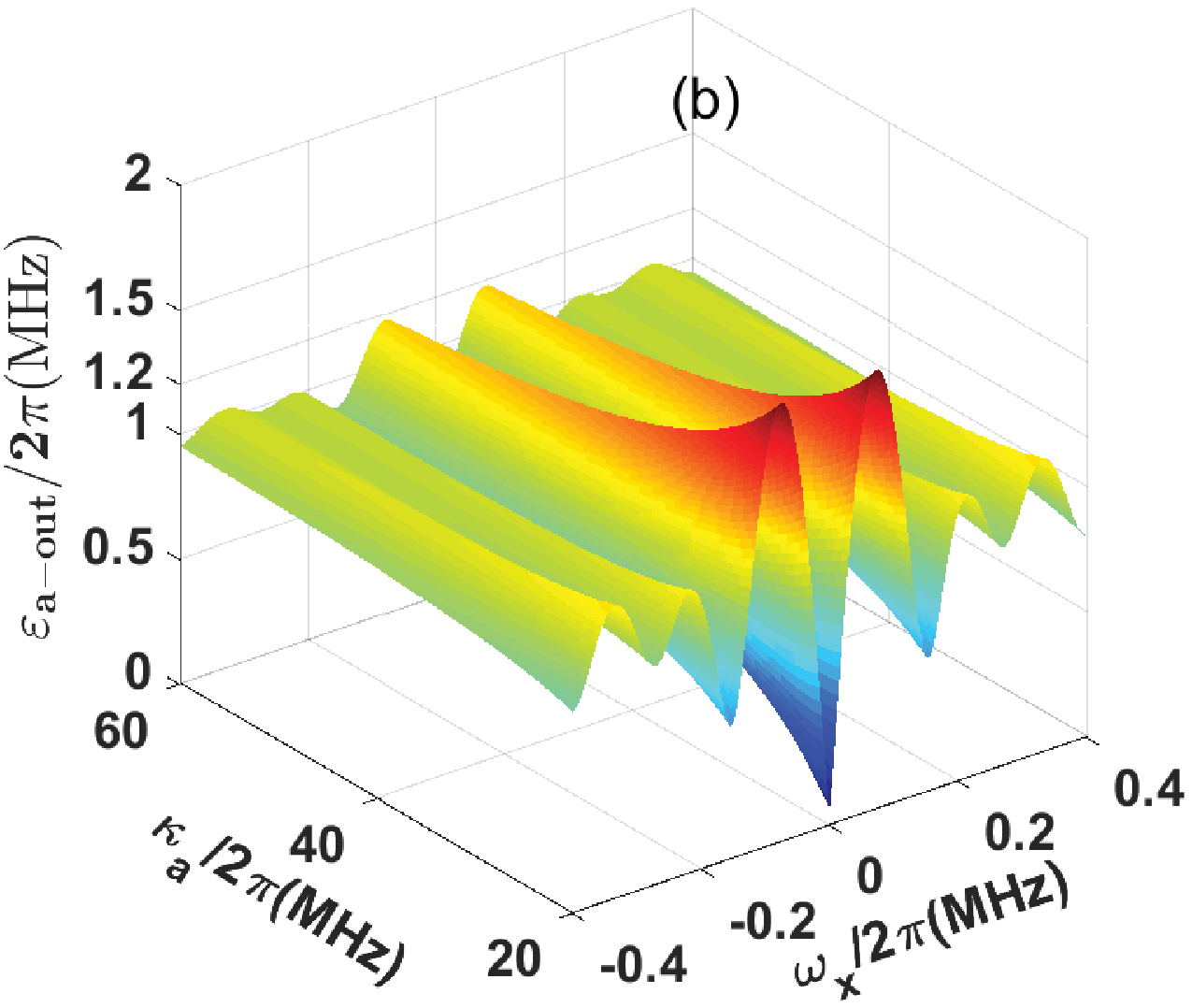}
\includegraphics[width=8cm]{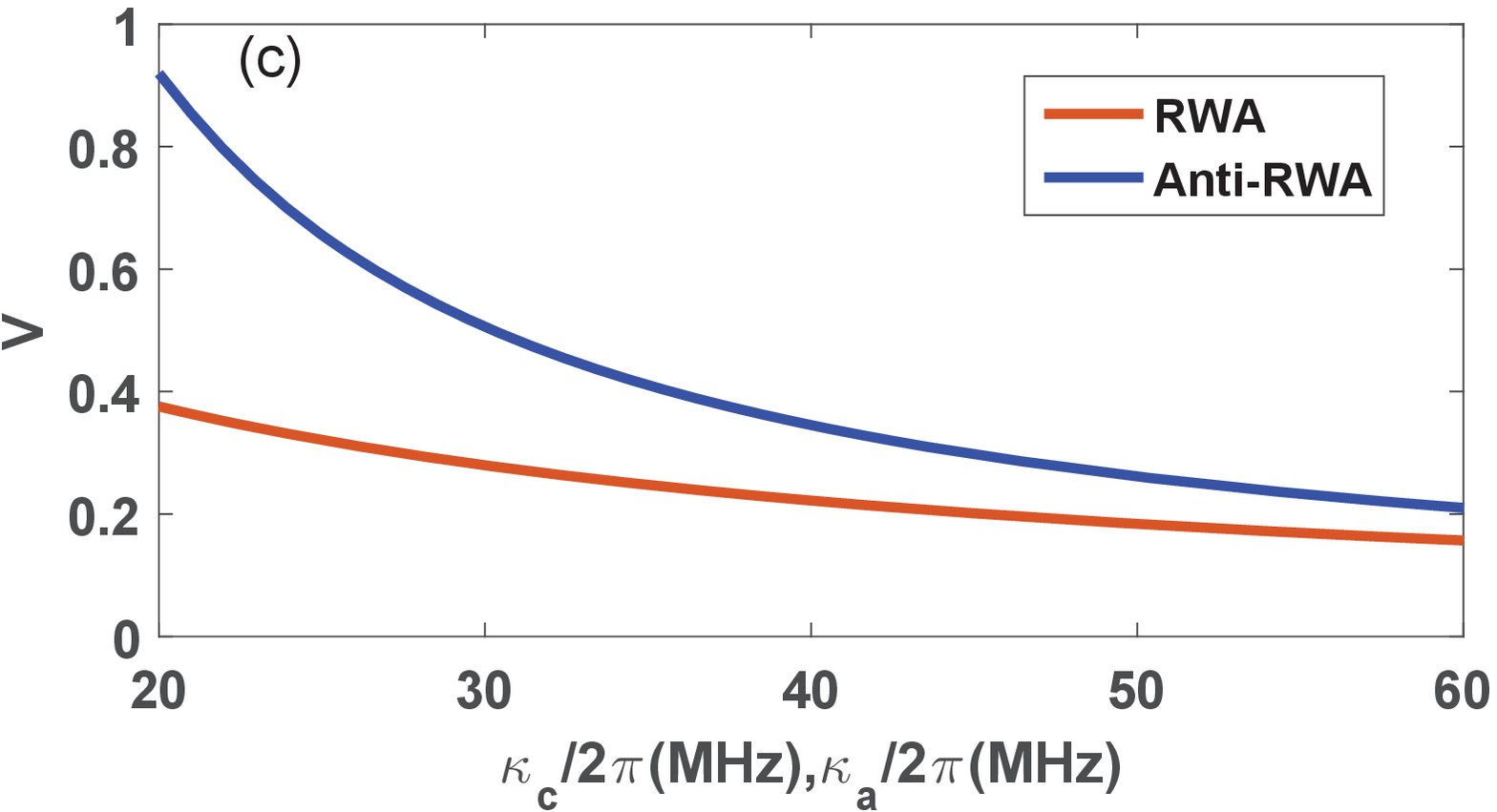}
\caption{(a) RI fringes for RWA with respect to the decay
rate $\protect\kappa _{c}$; (b) RI fringes for anti-RWA
with respect to the decay rate $\protect\kappa _{a}$. (c) Visibility $V=\frac{max(\varepsilon _{j-out})-min(\varepsilon _{j-out})}{max(\varepsilon _{j-out})+min(\varepsilon _{j-out})}$
corresponding to the cases of (a) and (b) versus the decay rate $\kappa_j$ with $j=c,a$ . Other parameters are
from Refs. \protect\cite{nphys-11-275,ncomm-6-6193,pra-90-053809}: $%
\protect\omega _{m}=42.3$ $\text{MHz}$, $\protect\varepsilon _{c}=$ $2\protect%
\pi \times 1.0$ $\text{MHz (RWA)}$, $\protect\varepsilon _{a}=$ $2\protect\pi %
\times 1.0$ $\text{MHz (anti-RWA)}$, $|G_{r}|=|G_{b}|=2\protect\pi \times 0.58$
 MHz, $\protect\gamma _{m}=2\protect\pi \times 20$ kHz, $\protect\omega _{lc}-%
\protect\omega _{la}\simeq \protect\omega _{m}$, $\protect\tau _{1}=4$ $\protect%
\mu$s, $T=4$ $\protect\mu$s, and $\protect\tau_{2}=0.1$ $\protect\mu$s.}
\label{fig34}
\end{figure}

We first simulate the RI fringes versus the detuning $\omega _{x}
$ with different operation time variables $\tau_{1}$, $T$, and $\tau _{2}$ in Fig. \ref{fig33}. Under the driving on mode $a$ ($c$) for the
RWA (anti-RWA), we find no (an) enhancement in the output field $\varepsilon _{c-out}$ ($\varepsilon _{a-out}$) for optical mode $c$ ($a$).
This phenomenon can be understood as follows. For the system working in the regime of RWA,
the interaction Hamiltonian $g[\alpha (t)bc^{\dag }+\alpha^{\ast}(t)b^{\dag }c]$ is just in a form of energy conservation. In contrast,
if the model works in the regime of anti-RWA, the interaction Hamiltonian $g(\zeta^{\ast}ab+\zeta a^{\dag}b^{\dag})$
works as a two-mode squeezing operator, which enhances the
RI fringes by the squeezing parameter $|G_{b}|=|g\zeta|$.
In such an open system, the energy for this enhancement is from the control field $\varepsilon_{c}$
via the SBS processes, where a high frequency optical
photon in mode $c$ is transformed into a low frequency photon in mode $%
a$ and an acoustic phonon $b$ as a Stokes process. Due to this reason,
enhancement of the RI fringes occurs within the interaction time duration
$\tau _{1,2}$, but not during the free evolution. When the interaction time $\tau_{1,2}$ is long enough,
the RI fringes for anti-RWA could be much larger than the ones for RWA.

\begin{figure*}[tbph]
\centering
\includegraphics[width=12cm]{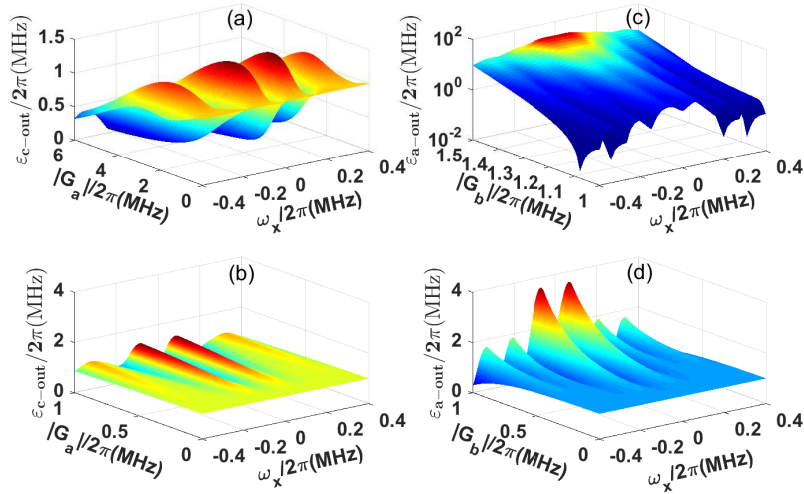}
\caption{(a,b) RI fringes for RWA  with respect to the
coupling strength $|G_{r}|$, where (b) is a zooming-in plot of (a). (c,d) RI fringes for anti-RWA
as a function of the coupling strength $|G_{b}|$, where (c) and (d) are complementary for clarity. Other
parameters are from Refs. \protect\cite{nphys-11-275,ncomm-6-6193,pra-90-053809}:
$\protect\omega _{m}=42.3$ MHz, $\protect\varepsilon_{c}=$ $2%
\protect\pi\times 1.0$ MHz (RWA), $\protect\varepsilon_{a}=$ $2\protect%
\pi\times 1$ MHz (anti-RWA), $\protect\kappa_{a}=\protect\kappa_{c}=2\protect\pi\times 30$ MHz,
$\protect\gamma_{m}=2\protect\pi\times 20$ kHz, $\protect\omega_{lc}-\protect\omega_{la}\simeq\protect\omega _{m}$,
$\protect\tau _{1}=4$ $\protect\mu$s, $T=4$ $\protect\mu$s, and $\protect\tau_{2}=0.1$ $\protect\mu$s.}
\label{fig35}
\end{figure*}

Our focus is now particularly on the anti-RWA case. As presented in Fig. \ref{fig34},
we show the RI fringes as functions of the detuning $\omega_{x}$ for
different decay rates. To show the fringes more clearly, the output fields $\varepsilon_{c-out}$ and $\varepsilon_{a-out}$ for
modes $c$ and $a$ are plotted in the same-scaled vertical axes. Compared with the output field $\varepsilon _{c-out}$
for mode $c$, the fringes of the output field $\varepsilon _{a-out}$  own an obviously better visibility [see  Fig. \ref{fig34} (c)] due to the properties of
the squeezing in the regime of anti-RWA. Moreover, with this enhancement, the RI
fringes of output field $\varepsilon _{a-out}$ for anti-RWA is more robust against the decay than the
counterparts of output field $\varepsilon _{c-out}$ for RWA. In our scheme, the contribution on the relative phase for
interferences is from two parts, one of which is from the free evolution of
the acoustic mode $b$ and the other of which is from the energy shift for the control
field. As a result, there is an fringe valley at $\omega_{x}\simeq\pm 2\pi\times
0.27$ MHz especially in the regime of anti-RWA.

To study the relationship between the effective coupling strength $|G_{r,b}|$
and the RI fringes, we have also calculated the RI fringes versus the detuning $%
\omega _{x}$ and the effective coupling strength $|G_{r,b}|$ in Fig. \ref{fig35}.
In the regime of RWA [Fig. \ref{fig35} (a,b)], when the effective coupling $|G_{r}|$ is very small, it is
hard to have an effective interaction between the
optical mode $c$ and the acoustic mode $b$. As such, no RI fringe appears. With the increase
of the effective coupling $|G_{r}|$, however, the above interaction could be somewhat enhanced, which yields maximum strength of RI
fringes at $|G_{r}|=2\pi \times 1.02$ MHz. In this case, the effective decay
rate $\Gamma _{r}$, increasing with $|G_{r}|$, is always for loss, which yields
disappearance of the RI fringes in the case of a very large coupling $|G_{r}|$.
For the anti-RWA case [Fig. \ref{fig35}(c,d)], the effective coupling strength $|G_{b}|$ with some values could greatly favor
the RI fringes. When the effective coupling strength is $|G_b|\leq 2\protect\pi\times0.557$ MHz,
the gain is just from the squeezing. But if $|G_b|$ is larger than $\protect\pi\times0.557$ MHz, the effective decay rate $\Gamma_b$
turns to be negative, which implies a gain. As a result, the enhancement of the RI fringes in this case is from both the squeezing and the
negative $\Gamma_b$. From our calculation, the best RI is achieved at $|G_b| \approx2\protect\pi\times1.0$ $\text{MHz}$ [see the dip in Fig. \ref{fig35}(c)].
But if this gain is too large, e.g., $|G_b|>2\protect\pi\times1.5$ MHz, the RI would be invisible.


\section{Conclusion}

In summary, we have presented two kinds of RI fringes with SBS in the RWA and
anti-RWA regimes, respectively, which can be switched via the
driving field on different optical modes. By comparison of the two fringes,
we show that the RI in the regime of anti-RWA could be benefited from squeezing property,
which enhances the visibility and strength of the fringes and robust to the decays.
Since the scheme is fully within the reach of current laboratory techniques, we believe
that our proposal provide a new way to building better Ramsey interferometers.

\begin{acknowledgments}
JQZ and XFQ thanks Prof. Yong Li and  Dr. Xiao-Feng Yi  for helpful discussions. This work is supported by MOST of
China under grant No. 2017YFA0304503, by National Natural Science Foundation
of China under grants No. 91636220, No. 91421111, No. 11674360 and No.
11504430.
\end{acknowledgments}

\end{document}